\documentclass[flushrt]{emulateapj}

\begin{document}

\title{Galaxy Clusters in Formation: Determining the Age of the Red-Sequence in Optical and X-ray Clusters at $\lowercase{z} \sim 1$ with HST\altaffilmark{1}}
\author{Benjamin P. Koester\altaffilmark{2,3}, Michael D. Gladders\altaffilmark{2,3,4}, David G. Gilbank\altaffilmark{5}, H.K.C. Yee\altaffilmark{6}, Kyle Barbary\altaffilmark{7,8}, Kyle S. Dawson\altaffilmark{7,9}, Joshua Meyers\altaffilmark{7,8}, Saul Perlmutter\altaffilmark{7,8}, David Rubin\altaffilmark{7,8}, Nao Suzuki\altaffilmark{7}}
\altaffiltext{1}{Based on observations made with the NASA/ESA Hubble Space Telescope and obtained from the data archive at the Space Telescope Institute. STSci is operated by the Association of Universities for Research in Astronomy, Inc. under the NASA contract NAS 5-26555. The observations are associated with program 10496}
\altaffiltext{2}{Department of Astronomy and Astrophysics, University of Chicago, Chicago IL 60637, USA}
\altaffiltext{3}{Kavli Institute for Cosmological Physics, The University of Chicago, Chicago IL 60637, USA}
\altaffiltext{4}{Visiting Associate, Observatories of the Carnegie Institution of Washington, Pasadena, CA 91101}
\altaffiltext{5}{Department of Physics and Astronomy, University of Waterloo, Waterloo ON, Canada, N2L 3GI}
\altaffiltext{6}{Department of Astronomy and Astrophysics, University of Toronto, Toronto, ON, Canada, M5S 3H4}
\altaffiltext{7}{E.O. Lawrence Berkeley National Lab, 1 Cyclotron Rd.,Berkeley,CA,94720}
\altaffiltext{8}{Department of Physics, University of California Berkeley,Berkeley,CA,94720}
\altaffiltext{9}{Department of Physics and Astronomy, University of Utah, Salt Lake City, UT,84112}
\begin{abstract}
Using deep two-band imaging from the \textit{Hubble Space Telescope},
we measure the color-magnitude relations (CMR) of E/S0 galaxies in a set of 9 optically-selected 
clusters principally from the Red-Sequence Cluster Survey (RCS) at $0.9 < z < 1.23$. 
We find that the mean scatter in 
the CMR in the observed frame of this set of clusters is $0.049 \pm 0.008$, as compared to 
$0.031 \pm 0.007$ in a similarly imaged and identically analyzed X-ray sample at similar redshifts. 
Single-burst stellar population models of the CMR scatter suggest 
that the E/S0 population in these RCS clusters truncated their star-formation at $z_l \simeq 1.6$, 
some 0.9 Gyrs later than their X-ray E/S0 counterparts which were truncated at $z_l \simeq 2.1$.
The notion that this is a manifestation of the differing evolutionary states of the two populations 
of cluster galaxies is supported by comparison of the fraction of bulge-dominated galaxies found in the two samples 
which shows that optically-selected clusters contain a smaller fraction of E/S0 galaxies at the their cores.

\end{abstract}
\section{Introduction}
The color-magnitude relation (CMR) in galaxies in the cores of galaxy clusters is a fossil record 
of the integrated star-formation histories of the constituent galaxies \citep[see][for a review]{renzini06}.
The slope of the CMR encodes the build-up of the mass-metallicity relation, while its
scatter is thought to reflect the variation in the formation epochs of the cluster 
galaxies \citep{kodama97,kauffmann98,bernardi05}. 

Mounting observational data has culminated in a picture of galaxy
formation that attempts to explain the conversion of star-forming ``blue-cloud'' 
galaxies into passively-evolving ``red-sequence'' galaxies \citep[e.g.,][]{faber07,mei09}. 
This picture cites some combination of the quenching of star formation in the blue-cloud and 
the merging of already red galaxies as driving the build-up of the red-sequence. 

The high resolution imaging delivered by HST/ACS plays an important role
in disentangling these effects, as it allows color measurements even for faint galaxies, 
and robust morphological 
measurements, which include either manual classification of E, S0, Sp, and Irr 
\citep[e.g.,][]{postman05} or automated separation of E/S0 types from later types 
\citep[e.g.,][]{abraham07}. 
In particular, combined color and morphological data can be used to 
measure the color scatter of 
the E/S0 galaxies in the CMR, which places constraints
on the formation epochs of stars in cluster galaxies \citep[e.g.,][]{bower92b}.

Until now, $z \sim 1$ systematic space-based studies of the CMR in 
the cores of galaxy clusters have been 
conducted almost exclusively using a handful of X-ray selected galaxy clusters 
\citep{vandokkum98,vandokkum00,blakeslee03,mei06,blakeslee06,mei09}. The consensus picture
from these studies is that stars in these galaxies were formed at 
$z \gtrsim 2$. 

In this Letter, we present the first precision statistical
measurements of the scatter of the CMR in an optically-selected sample
of galaxy clusters at $z\sim1$. Where necessary, we adopt a flat,
$\Omega_m=0.30$ cosmology with $H_0=70$ km s$^{-1}$ Mpc$^{-1}$.

\section{Data}
Imaging for this study was acquired as part of the \emph{HST} Cluster SN Survey 
Program (Program Number 10496, PI: Perlmutter). 
Nine optically-selected clusters were imaged at various telescope roll angles for a total of 
5000-16000s with the 
Advanced Camera for Surveys (ACS) on multiple visits (375s-500s per sub-exposure)
in $z_{850}$ (F850LP) 
and for 1000-8000s in $i_{775}$ (F775W) \citep{dawson09}. 
Eight clusters are derived from the Red-Sequence Cluster 
Survey \citep[RCS,][]{gladders05}, and the ninth, XLSS J0223.0+0436, was originally 
included in this sample as an IR-detected cluster from the SpARCS survey \citep{muzzin08,wilson08}, but
had previously been detected in both the IR \citep{andreon05} and X-ray \citep{pierre04}.
We consider it here as an optical-IR detected cluster. With the exception RCSJ J0337.8-2844 
an extensive spectroscopic program has
obtained 5-30 confirmed spectroscopic members per cluster, enough to compute a velocity
dispersion in many cases \citep{gilbank08,hicks08,gilbank09}. Three of the X-ray selected 
clusters, analyzed most recently in \citet{mei09} are taken from
\citep{rosati98}, and the WARPS cluster was first presented in \citep{perlman02}.

Images for each cluster were coadded using Multidrizzle
\citep{koek02}, with a Gaussian kernel, $\tt{pixfrac=0.8}$, and a final
output pixel size of 0.03 arcsec/pixel.  The detailed photometric and
morphological pipelines developed for this work are described and
tested in detail in \citet{koester09}.  In brief, object detection was
performed with SExtractor \citep{bertin96}, and the resulting catalogs
were manually cleaned of diffraction spikes, pupil ghosts, and
saturated stars. The final galaxy catalog was distilled from the full
catalog following \cite{leauthaud07}.
Total magnitudes are taken as $\tt{MAG\_AUTO}$ magnitudes with a correction applied
\citep[e.g.,,][]{bernstein02a,bernstein02b, benitez04,blakeslee06} appropriate for E galaxies.
As noted in \citet{blakeslee03}, the differential smearing of the PSF 
between $i_{775}$ and $z_{850}$ makes the PSF approximately $10\%$ broader in $z_{850}$. 
We correct this using a PSF matching
scheme that places both the $i_{775}$ and $z_{850}$ images on the same footing 
by convolving the latter image with the PSF of the former, and vice-versa.

The PSF model was built in each bandpass using the SExtractor $\tt{MU\_MAX-MAG\_AUTO}$ parameters 
to select acceptable PSF stars in the coadded images, which were then visually checked. 
In a circularly symmetric $\sim 1'$ radius, a single PSF model adequately describes the aggregate
distortion realized in the final image due to HST/ACS optics and temporal 
variations thereof \citep{jee07,rhodes07}, charge diffusion \citep{krist03},
the so-called red-halo effect \citep{sirianni05,jee07}, and the drizzling process. 
For the precision photometry required in this study, this 
necessarily restricts us to considering the inner $500h^{-1}$ kpc of each 
cluster, where the PSF is approximately constant, and the convolutions required
for PSF-matching are minimally position dependent. 

In the coadded and convolved $z_{850}$ image of each cluster, we
construct adaptively-determined apertures with $\tt{GALFIT}$
\citep{peng02} by fitting Sersic profiles constrained to $1 \le n \le
4$, thus returning an estimate of an effective radius, $R_e$ for each
object. Nearby objects ($\le 3R_e$) are simultaneously fit, and all
remaining objects within $6R_e$ are masked, and the sky estimated as
in \citet{haussler07}.  The same aperture is applied in the $i_{775}$
band to measure the $i_{775}-z_{850}$ color.  Errors on the aperture
magnitudes are determined for each galaxy by comparing the repeat
observations of objects as a function of magnitude. Typical errors
range from 0.02 mags at $z_{850}=21$ to 0.05 magnitudes at
$z_{850}=24$.

We use an automated scheme, $\tt{MORPHEUS}$ \citep{abraham07}
to separate E/S0 galaxies from the irregulars and spirals, taking objects with
a Gini coefficient $>0.45$ and asymmetry parameter $<0.1$ to be E/S0 galaxies. 
Because $z_{850}$ stays redward of the 4000\AA~ break until $z \sim 1.3$, 
the signal-to-noise ratio (S/N)
 for early-type galaxies remains considerable. 
Based on comparison to the manual classifications for RDCS
J1252.9-2927 in \citet{postman05}, the E/S0 galaxies we classify at $z
= 1.24$ comprise a $90\%$ pure and complete sample. 

\section{Measuring the Color-Magnitude Relation}
Color-magnitude diagrams were created for each cluster, and the red-sequence color limits were 
determined by eye. Objects falling within
these broad limits ($\simeq 3 \sigma$ of the final best-fit line) with luminosities
greater than $0.3L_*$ ($z_{850}=24$ at $z=1.24$) were taken to be the red-sequence. 
The luminosity limit is strictly chosen to be bright enough to ensure the morphological
classifications are robust. A line was fit to E/S0 objects within these
bounds using a simple least-squares method, and was sigma-clipped at 3 $\sigma$ in $i-z$ color.
This was iterated until convergence to derive the slope and zero point of the
best-fit line, and the scatter of E/S0 objects (including the rejected outliers)
about this line was computed using a bi-weight estimator of scale \citep{beers90}.

The best-fit slopes of two clusters, RCS J2156.7-0448 and RCS J2345.4-3632, 
are not robust to the color cuts used, due largely to the relative paucity of E/S0 galaxies
at the faint end of their CMRs ($z_{850}> 22.5$). 
We thus fix the slope for these clusters at $-0.05$ and the reported
scatter is measured about this line; the resulting scatter about the
best-fit line varies by only $\sim 0.015$ mags depending on the
line used.

To assess uncertainties, we run 1000 bootstrap resamples
on each cluster, and compute the mean and standard error of the red
sequence fits to these resamples. There is a further contribution
to scatter due to photometric error. Using the method of
\citet{stanford98} we find that it
contributes no more than 0.01-0.02 magnitudes in quadrature to the
measured scatter; we correct the measured scatters accordingly.

Finally, we note that the derived CMR is
measured inside a fixed 0.5 $h^{-1}$ Mpc cluster-centric radius, 
the approximate observed field-of-view of most of these
observations.  We do not attempt to compute $R_{200}$ or any other
scale radius, which would require additional dynamical data, X-ray
data, or possibly wider field images \citep[e.g.,][]{rozo08}.

\section{The CMR in $\lowercase{z} \sim 1$ RCS Clusters}
Table 1 shows the measured CMRs for our 9 optically-selected clusters and 
also our own measurements of a subset of archival X-ray selected clusters also
measured in $i_{775}$ and $z_{850}$. The Table also displays $N_{E/S0}$, the number 
of E/S0 galaxies used in measuring the CMR, and a richness $N_{0.5}$, 
which is the number of red-sequence galaxies brighter than $0.3L*$, within $0.5 h^{-1}$ 
Mpc of the BCG. RCS J0337.8-2844 and RCS J2156.7-0448, the two 
least rich clusters, exhibit the largest scatter.
RCS J2319.8+0038 and RCS J0439.6-2904, which are both
X-ray detected \citep{hicks08,gilbank08}, and the X-ray selected clusters
have relatively high $N_{0.5}$ richnesses.

In this Letter, we focus on the distribution of scatters. In the observed
frame, it is apparent that the optical and X-ray samples have discordant
average scatters in observed $i_{775}-z_{850}$: the optical sample, at $0.049 \pm 0.008$ is 
nearly 50\% larger than the X-ray scatter at $0.031 \pm 0.007$. A $t$-test of 
the means of the two samples rejects the null hypothesis that the two 
samples come from the same underlying population at the 95\% level.  

The scatter is thought to reflect variations in galaxy formation time 
\citep[but see][]{menci08}. It has thus been modeled using Bruzual-Charlot
single-burst models \citep{bruzual03}, which posit that all the stars 
in a given galaxy are formed in a coeval burst at some redshift, $z_f$,
and evolve passively thereafter. The scatter in the CMR results from the variation
in formation times for the constituent galaxies that have formed in single bursts between
reionization, $z_r$, and some lower redshift, $z_l$. 

\begin{figure}[htp]
\centering
\includegraphics[scale=0.38,angle=90]{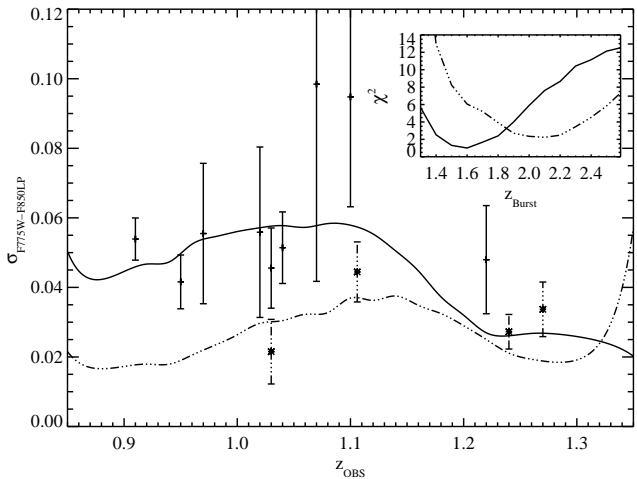}
\caption{Bruzual-Charlot Single-Burst Models of the Scatter. Given the
cluster redshift, $z_{obs}$, stars with
error bars represent the scatter of the red-sequence in X-ray selected clusters, while
the normal crosses and errors show the scatters for the optically-selected clusters.
The best-fit Bruzual-Charlot single-burst models for the X-ray and optical samples
are shown as dashed and solid lines, respectively. Inset: $\chi^2$ minimization
for the X-ray and optical models, indicating the lowest redshift, $z_{burst}$, of the 
epoch during which galaxies formed. The two samples clearly favor different minima in the context
of this model. Alternatively, taking $z_{burst}=1.6$ in the x-ray sample, the $\chi^2=6.1$, 
a result that is discordant with the optically-selected best-fit $z_{burst}=2.1$ model 
at the few sigma level.  }
\end{figure}

Using the ACS $i_{775}$ and $z_{850}$ response curves as described 
in \citet{sirianni05}, we determine the expected observed scatter 
in $i_{775} - z_{850}$ using a single-burst spectrum, following the Monte
Carlo method first described in \citet{vandokkum98}. Assuming a galaxy formed
at some redshift, $z_f$, we create 10,000
``galaxies'' by drawing from a random uniform formation age distribution corresponding
to a redshift range of $z_l \le z_f \le z_r$.
For each $z_l$, the last redshift at which galaxies formed, the model predicts 
a scatter in $i_{775}-z_{850}$ that would
be observed at some redshift, $z_{obs}$. For a given set of clusters, 
we then choose the best-fit model by minimizing $\chi^2$:
\begin{equation}
\chi^2=\Sigma\frac{(\sigma^{mod}_{iz}-\sigma^{obs}_{iz})^2}{\sigma_{err}^2.},
\end{equation}
where the ``mod'' and ``obs'' superscripts refer to the predicted model 
and observed data scatters, and $\sigma_{err}$ is the error on each observed
$\sigma^{obs}$. In Figure 1, we show the result of this process for both 
the optical (solid) and X-ray (dashed) samples. 
The optical
sample favors $z_l= 1.6$, and the X-ray sample favors
$z_l= 2.1$, suggesting that the E/S0 galaxies on the red-sequence in 
X-ray selected clusters ended their star formation some 0.9 Gyr earlier.

\section{Discussion}
The small number of E/S0 galaxies, $N_{E/S0}$, in some of the less rich clusters
highlights a possible concern.  Simple Monte Carlo simulations of
random draws from a normal distribution indicate that at $N=10$, even
the bi-weight estimator of scale returns a scatter that is biased high
by 10\%; this is reduced to 3\% by $N=20$, and $ \lesssim 1\%$ at
N=40. Contamination by non-cluster members also contributes to
scatter.  Randomly adding a single galaxy that passes the red-sequence
color cuts adds as much as an additional 10\% bias at $N=10$.
Fortunately, our analysis of the GOODS\footnotemark[9]
\citep{giavalisco04} fields indicate that the average background (for
morphologically selected galaxies passing our color and luminosity
cuts) is $\lesssim 1$ galaxy.

Another concern is line-of-sight projection of large structures.  RCS
J0439.6-2904 is known to show two redshift peaks at $z=0.9435$ and
$z=0.9681$ \citep{gilbank07,cain08}, for a separation of $\simeq 7000$
km s$^{-1}$. A cursory analysis of the halo catalogs from the VIRGO
simulations \citep{evrard02} shows that halo-halo projections of this
sort are expected in about $\simeq 10-20\%$ of RCS clusters in this
mass range at $z \sim 1$; a similar value is found in spectroscopic data
\citep{gilbank07}. At such a separation, Monte Carlo simulations show
that two projected halos each with $\sigma_v = 1000$ km s$^{-1}$ have
a resultant measured scatter inflated by only $0.003$ magnitudes,
which has no significant effect on our results.

Spectroscopy does not yet exist for RCS J0337.8-2844 and 
spectroscopic confirmation is pending.
Only limited spectroscopy exists for RCS J2156.7-0448 (5 confirmed
spectroscopic members), and its measured X-ray flux is consistent with
zero \citep{hicks08}, although the strong lensing in this system
\citep{gladders03}
indicates that its mass is cluster scale.
Regardless, we find that removing one or both of these clusters does
not affect our conclusions. The other seven clusters are robustly
spectroscopically confirmed \citep{gilbank09}.

\footnotetext[9]{Based on observations made with the NASA/ESA Hubble
Space Telescope, obtained from the data archive at the Space Telescope
Institute. STSci is operated by the Association of Universities for
Research in Astronomy, Inc. under the NASA contract NAS 5-26555. The
observations are associated with program 9425.}

For three of the X-ray clusters (all except WARP J1415.1+3612),
\citet{mei09} place constraints on $z_l$ with single-burst models,
using a combination of HST data, manual morphological measurements,
and spectroscopic confirmation where available. They generally find
formation epochs of $z_l \simeq 2.2$, only slightly higher than our
own measurements. As part of the validation process of our pipeline,
we compare our measured CMR parameters in these 3 X-ray clusters with
previous work, and find that they agree within errors.

\begin{figure}[htp]
\centering
\includegraphics[scale=0.35,angle=90]{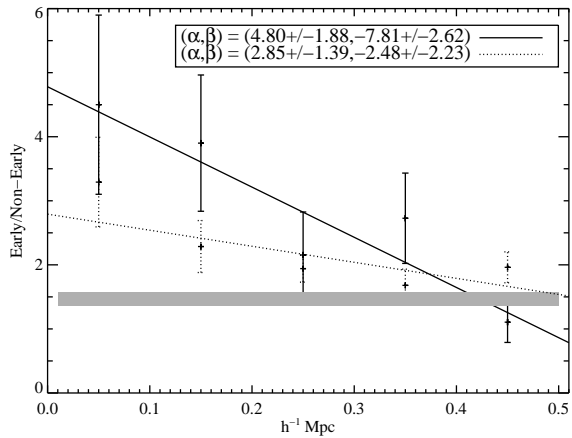}
\caption{Radial Dependence of E/S0 Fraction. X-ray selected clusters (solid)
show a larger fraction of color-selected E/S0 galaxies than optically-selected 
(dotted) clusters, and both fractions approximately converge to the GOODS fields 
values (gray bar) by 0.5 $h^{-1}$ Mpc. Best-fit regression lines are overplotted
as well, and the intercept and slope $(\alpha,\beta)$ are shown in the legend.
The X-ray fraction at $r \le 0.4 h^{-1}$ Mpc is consistent with that measured
in \citep{postman05}.}
\end{figure}

For the 4 optical clusters in this study that have
velocity dispersion estimates, we find the distribution of velocity dispersions
is similar to the X-ray sample, with the RCS clusters having
$670 \lesssim \sigma_v \lesssim 1080$ km s$^{-1}$, and the X-ray sample 
$650 \lesssim \sigma_v \lesssim 747$ km s$^{-1}$. 
The number of E/SO galaxies (Table 1) suggests that the clusters in the
X-ray sample are as rich as the richest optical clusters, and a
$t$-test reveals no significant difference between richnesses of the
two samples.  Additionally, the fact that we have run the same
automated pipeline on all the clusters rules out
systematic differences that might be expected in measurements
conducted in different studies, or in different bandpasses.

The observed difference in red-sequence scatters between the two
samples therefore suggests that the galaxy populations in the clusters
themselves are different.
To test this further, we measure 
the composite red-sequence bulge-dominated (E/S0) fraction in each sample 
as a function of cluster-centric radius, shown in Figure 2. Both samples reach the GOODS
field value by $0.5h^{-1}$ Mpc.
Overplotted are best-fit regression lines estimated using
$\tt{linmix\_err}$ \citep{kelly07}, which both exhibit a significant
correlation of the fraction with radius. The regression parameters
themselves (Figure 2, legend) disagree at the $\simeq 2\sigma$
level. This suggests that E/S0 galaxies form a larger fraction of the
cores of X-ray clusters, and fall off faster than the optical clusters
before reaching the field value. At lower redshifts the development of
the morphology-density relation, particularly through the build-up of
S0 galaxies on the red sequence, appears correlated with cluster age
\citep{dressler97}, and so these differences also suggest that the
optical clusters are younger. We note however that the radii in Figure
2 are metric, and it is not clear that these trends would be mitigated
by the use of an aperture scaled to the virial radius. The similarity
of the X-ray and optical velocity dispersions is, however, evidence
that the virial radii are similar.

The measurements presented in this study have consequences for studies
of cluster evolution and galaxy formation. Distant optically-selected
clusters likely represent a younger state of massive
halos compared to their X-ray selected counterparts. If cluster
galaxies from the most highly-biased regions of the Universe formed
earlier in the X-ray sample, as suggested by our analysis, one might
be tempted to conclude that the clusters themselves collapsed earlier,
and that the presence of a hot ICM reflects this more evolved
state. In this scenario, X-ray selected clusters only represent a
subset of the most massive halos at $z \sim 1$, and optical samples
fill out the part of the mass function that is younger, and not X-ray
bright. However, absent X-ray data for the full optical sample, it is
difficult to draw any definite conclusions on the relationship of
cluster dynamical evolution to X-ray luminosity and the red-sequence.

\acknowledgements{We thank Keren Sharon for a critical reading of the manuscript, and Gus Evrard and Jiangang Hao for helpful discussions. Financial support for this work was provided by NASA through program GO-10496 from the Space Telescope Science Institute, which is operated by AURA, Inc., under
NASA contract NAS 5-26555. This work was also supported in part by the Director, Office
of Science, Office of High Energy and Nuclear Physics, of the U.S. Department of Energy
under Contract No. AC02-05CH11231.}

\begin{table*}[h] \caption{Clusters and CMR Parameters Used in this Study}
\begin{center}
\begin{tabular}{l c c c c c c c}
\hline
Name \tablenotemark{1}& $N_{E/S0}$ & $z_{spec}$ & $(i-z)$ slope & $(i-z)$ scatter & $N_{0.5}$ \tablenotemark{2}\\
\hline
RCS J2319.8+0038     & $39$ & $ 0.91$ & $-0.066 \pm 0.012$ & $ 0.054 \pm 0.006$ & $57.6 \pm  1.3$\\
RCS J0439.6-2904     & $23$ & $ 0.95$ & $-0.053 \pm 0.019$ & $ 0.042 \pm 0.008$ & $23.3 \pm  0.9$\\
RCS J1511.0+0903     & $12$ & $ 0.97$ & $-0.035 \pm 0.053$ & $ 0.055 \pm 0.020$ & $ 9.4 \pm  1.0$\\
RCS J0221.4-0321     & $18$ & $ 1.02$ & $-0.042 \pm 0.073$ & $ 0.056 \pm 0.025$ & $ 26.9 \pm  1.3$\\
RCS J0220.9-0333     & $15$ & $ 1.03$ & $-0.010 \pm 0.022$ & $ 0.046 \pm 0.012$ & $14.9 \pm  1.4$\\
WARP J1415.1+3612  \tablenotemark{3}& $12$ & $ 1.03$ & $-0.018 \pm 0.020$ & $ 0.022 \pm 0.009$ & $30.8 \pm  0.9$\\
RCS J2345.4-3632 \tablenotemark{4}    & $16$ & $ 1.04$ & $-0.050 \pm 0.000$ & $ 0.051 \pm 0.011$ & $24.5 \pm  1.5$\\
RCS J2156.7-0448 \tablenotemark{4}& $7$ & $ 1.07$ & $-0.050 \pm 0.000$ & $ 0.098 \pm 0.041$ & $ 13.6 \pm  1.4$\\
RCS J0337.8-2844 \tablenotemark{5}    & $11$ & $ 1.10$ & $-0.108 \pm 0.090$ & $ 0.095 \pm 0.032$ & $ 18.6 \pm  1.3$\\
RDCS 0910+5422 \tablenotemark{6} & $27$ & $ 1.11$ & $-0.044 \pm 0.013$ & $ 0.044 \pm 0.009$ & $21.5 \pm  0.8$\\
XLSS J0223.0+0436    & $18$ & $ 1.22$ & $-0.043 \pm 0.022$ & $ 0.048 \pm 0.016$ & $13.0 \pm  1.1$\\
RDCS J1252.9-2927 \tablenotemark{6} & $19$ & $ 1.24$ & $-0.038 \pm 0.011$ & $ 0.027 \pm 0.005$ & $21.5 \pm  0.7$\\
RDCS J0848.9+4452 \tablenotemark{6} & $23$ & $ 1.27$ & $-0.041 \pm 0.009$ & $ 0.034 \pm 0.008$ & $19.4 \pm  0.5$\\

\hline
\end{tabular}
\end{center}
$^1$ Number of E/S0 galaxies ($N_{E/S0}$), cluster spectroscopic redshifts ($z_{spec}$), observed $i_{775}-z_{850}$ CMR parameters.\\
$^2$ $N_{0.5}$: The GOODS background-subtracted number of galaxies within 2$\sigma$ of the best-fit line ($\sigma$ is our reported scatter) brighter than 0.3$L_*$ at the cluster redshift, and within $0.5h^{-1}$ Mpc.\\
$^3$ WARPS1415 X-ray detected cluster \citep{perlman02}.\\
$^4$ A fixed slope is used to measure the scatter, the best-fit slope is sensitive to the color cuts used. See text.\\
$^5$ Spectroscopic confirmation pending.\\
$^6$ See Mei et al., 2009\\
\label{table:cmdtable}
\end{table*}

\end{document}